\documentclass[structabstract]{aa}  
\usepackage{graphicx}
\usepackage{txfonts}
\usepackage{natbib}
\begin{document}
   \title{Evolution and CNO yields of $Z=10^{-5}$ stars and possible effects on CEMP production.}


   \author{P. Gil--Pons,
          \inst{1} 
          \inst{2}
          C. L. Doherty, \inst{2}
          H. Lau \inst{3}
	  S.W. Campbell, \inst{2}
          T. Suda, \inst{4}
          S. Guilani, \inst{1}
	  J. Guti\'errez,\inst{1}
          \and
          J.C. Lattanzio \inst{2}
          }

   \institute{Universitat Politecnica de Catalunya, Av. Canal Olimpic, 
          08840 Castelldefels, Barcelona, Spain
          \email{pilar@fa.upc.edu}
          \and
          Monash Centre for Astrophysics, Monash University, VIC 3800, Australia
          \and
          Argelander-Institut fuer Astronomie, Universit Bonn Auf dem Huegel 71, D-53121 Bonn, Germany
	  \and
          National Astronomical Observatory of Japan, Osawa 2-21-1, Mitaka, Tokyo 181-8588, Japan
             }

   \date{Received ; accepted }

\abstract{}{}{}{}{}
 
  \abstract
    {}   
{ Our main goals are to get a deeper insight
    into the evolution and final fates of intermediate-mass, extremely
    metal-poor (EMP) stars. We also aim to investigate the C, N, and O
    yields of these stars.}
      {Using the Monash University Stellar Evolution code
      MONSTAR we computed and analysed the
      evolution of stars of metallicity $Z = {\rm 10^{-5}}$ and masses between 4 and 9 
       $ M_{\odot}$, from their main sequence until the late thermally pulsing (super) 
      asymptotic giant branch, TP-(S)AGB
      phase.}
{Our model stars experience a strong C, N,
and O envelope enrichment either due to the second dredge-up
  process, the dredge-out phenomenon, or the third dredge-up early during the TP-(S)AGB phase.
Their late evolution is therefore similar to that of higher metallicity
objects. When using a standard prescription for the mass loss rates during the
TP-(S)AGB phase, the computed stars are able to lose most of their
envelopes before their cores reach the Chandrasekhar mass (${m_{Ch}}$), so our
standard models do not predict the occurrence of SNI${\rm 1/2}$ for
$Z = {\rm 10^{-5}}$ stars.  However, we find that the reduction of only one
order of magnitude in the mass-loss rates, which are particularly uncertain
at this metallicity, would prevent the complete ejection of the
  envelope, allowing the stars to either explode as an SNI${\rm 1/2}$ or
  become an electron-capture SN. Our calculations stop due to an
  instability near the base of the convective envelope that hampers
  further convergence and leaves remnant envelope masses between 0.25 ${
    M_{\odot}}$ for our 4 ${ M_{\odot}}$ model and 1.5 ${ M_{\odot}}$
  for our 9 ${ M_{\odot}}$ model. We present two sets of C, N, and O
  yields derived from our full calculations and computed under two
  different assumptions, namely, that the instability causes a practically
  instant loss of the remnant envelope or that the stars recover and
  proceed with further thermal pulses.}
   {Our results have implications for the early chemical evolution of the Universe and might provide 
   another piece for the puzzle of the carbon-enhanced EMP star problem.}

  \titlerunning{Evolution and CNO yields of intermediate mass $Z = {\rm 10^{-5}}$ stars}
  \authorrunning{Gil-Pons et al.}

  \maketitle

%

\section{Introduction}

   The initial mass function (IMF), the evolution, the final fates, and the
   yields of metal-free and metal-poor stars are keys to understanding the
   chemical evolution of the primitive Universe.

   The primordial IMF plays a central role in determining the different
   stellar contributions to the chemical enrichment processes in early
   galaxies. In particular, asymptotic giant branch (AGB) stars have a potential importance in the
   enhancement of carbon in observed metal-poor stars in the Galactic halo
   because they can contribute to formation of carbon-enhanced metal-poor
   (CEMP) stars through binary mass transfer and/or AGB winds.  The
   primordial IMF has been studied through numerical simulations of the
   collapse of gas clouds, whose composition is that expected from Big-Bang
   nucleosynthesis \citep{bro03,nak01}, and through the comparison of
   the results of population synthesis models with observations
   \citep{sud12a,sud13,iza09,kom07}. The IMFs preferred by different
   authors, even when using analogous approaches to the problem, differ
   widely.  For instance, \citet{bro03} conclude that the primordial IMF
   has to be biased towards very high masses ($\ga$ 100
   ${M_{\odot}}$), whereas the work by \citet{nak01} suggests there
   is a bimodal primordial IMF, peaking both at about 1
   ${M_{\odot}}$ and at about 10 ${M_{\odot}}$. \citet{iza09}
   conclude that a low-mass-dominated IMF is needed, whereas Suda et
   al. (2013) show the necessity of taking a primitive IMF
   biased towards higher masses into account.  All these results, even when
   contradictory, provide context and information on the problem of the IMF
   of the early Universe and the evolution and yields of the oldest
   stars in the Universe.

   From the point of view of stellar evolution, a huge amount of work has
   been devoted to the study of primordial to extremely metal-poor (EMP)
   stars. According to \citet{chr05}, these are stars with
   $\rm{[Fe/H]<-3}$. Earlier work on these stars has been done by
   \citet{dan82},
   \citet{fuj90}, \citet{cas97}, \citet{chi01}, \citet{sie02},
   \citet{mar01}, \citet{sud10}, \citet{gil05}, \citet{lau07},
   \citet{gil07}, \citet{cam08}, \citet{lau09}, \citet{sta10}, and
   \citet{lug12}.  These authors mostly studied the low- and
   intermediate-mass ranges, whereas \citet{heg00}, \citet{heg02},
   \citet{lim05}, \citet{mey06}, and \citet{hir08} extended evolutionary
   calculations to the high-mass range.  Still, important uncertainties in
   the input physics remain an enormous problem, even to the point where we
   cannot confidently predict the final fates (and consequently the
   yields) of stars of metallicity $Z \la 10^{-3}$ yet
   \citep{heg03}. 

   The main sources of uncertainties are the mass-loss rates
   due to stellar winds and the chemical and energy transport
   processes. These uncertainties hamper determination of the evolution
   of the core and envelope masses and, therefore, our knowledge of the
   final fates of intermediate-mass primitive stars as a function of
   initial mass, either as white dwarfs or as supernovae of different
   types.  In the cases of stars of solar or moderately low metallicity,
   observations can help us constrain the problem of the wind mass-loss
   rates and, up to a certain point, the chemical and energy transport in
   stellar interiors. On the other hand, most (if not all) of the primordial
   stars died long ago, and even though the number of observed very
   metal-poor stars is increasing thanks to large studies, such as the
   Hamburg/ESO \citep{wis00,chr01} survey, HK \citep{bee92} survey, and
   SEGUE(SDSS) \citep{yan09}, it is still difficult to get a fully coherent
   picture of the evolution between primordial and extremely low
   metallicities. 

 Chemical evolution models of the Galaxy \citep{kob06,ces06,chi05} are also
 useful for understanding the early Universe, since they provide
 information on which proposed stellar objects should be retained or
 discarded, such as SNI${\rm1/2}$. They can also be used for probing the
 primitive IMF \citep{sud13}. 
   
  \begin{figure}
   \centering
   \vspace{0.2cm}
   \includegraphics[scale=0.60]{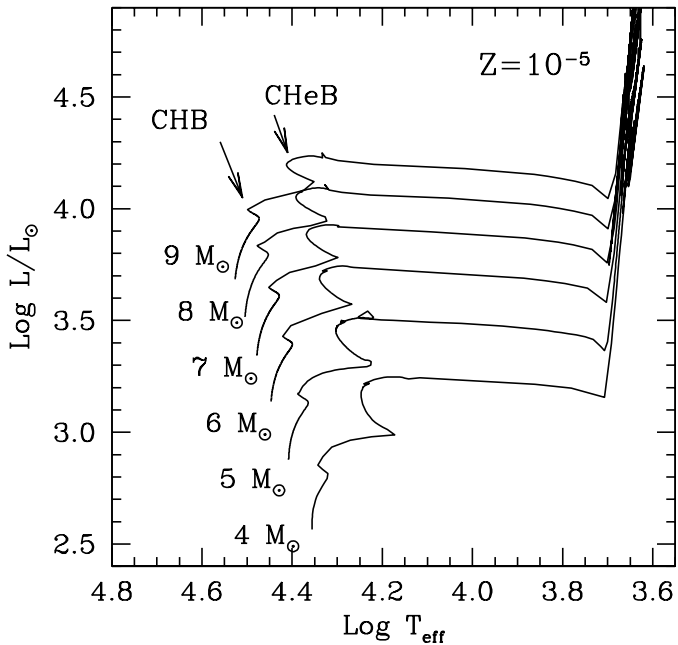}
   \caption{HR diagram of the $Z=10^{-5}$ models for masses between 4 and
            $9 M_{\odot}$.}
              \label{fig:hrs}
    \end{figure}

   Stellar models show that the physical properties of primordial to EMP
   stars, namely their higher overall compactness and smoother entropy
   profiles near the active burning shells compared to models of metal-rich
   stars \citep{fuj00}, cause peculiar nucleosynthetic and mixing
   behaviours.  For instance \citet{fuj90}, \citet{fuj00} and, more
   recently, \citet{sud04} and \citet{cam08} showed that the convective
   instabilities due to helium-core and shell flashes in low mass and low
   metallicity stars can extend to hydrogen-rich regions and cause hydrogen
   flashes which then enable the transport of carbon to the stellar
   surface. \citet{chi01} described carbon ingestion, a similar process
   occurring during thermal pulses of more massive thermally-pulsing asymptotic
   giant-branch, TP-AGB stars (${M_{ZAMS} \la 5}$ ${M_{\odot}}$).
 
   In this work we compute and analyse the evolution of stars having a
   metallicity of $Z = {\rm 10^{-5}}$ in a mass range between 4 and 9 ${M_{\odot}}$. 
   We follow the stellar evolution from the main sequence
   until the late stages of their TP-(S)AGB \footnote{We use this acronym to
   denote both AGB and Super-AGB stars.} phase and study in detail
   whether these objects show unusual mixing events like the cases
   mentioned above.
Our results represent progress with respect to former works on stellar
evolution at this metallicity and mass range. Instead of evolving only the
early stages of the TP-(S)AGB phase and then extrapolating the mass loss
and core growth to the end of the star's life \citep{gil07a}, we now
complete the computation of the stellar sequences almost until the end of
the TP-(S)AGB. The calculations only halt when an Fe-opacity-induced
instability of the type described in \citet{lau12} is encountered.  This
way we expect to get a better approach to the problem of the final fates
and CNO yields of the studied objects.
 
The present work is organised as follows. The second section is dedicated
to a brief description of the evolutionary code and the particular input
physics that are relevant to this work. The third section covers the
evolution prior to the TP-(S)AGB, with special attention to the corrosive
second dredge-up and dredge-out phenomena. The fourth section analyses
the evolution along the TP-(S)AGB.  The fifth section deals with the yields
and the final fates of our model stars.  In the last section we draw the
main conclusions that can be derived form our work.

\begin{table*}
\caption{
Relevant times, structure and composition values that characterise the 
evolution prior to the TP-(S)AGB.
}
\begin{center}
\begin{tabular}{cccccccc}
\hline
{${M_{ZAMS}}$}
  & {${t_{CHB}}$}
  & {${M_{He}}$}
  & {${t_{CHeB}}$}
  & {${M_{CO}}$}
  & {${X_c(C)}$}
  & {${X_c(O)}$}
  & {$\rm{C-burning}$}\\
 $M_{\odot}$ 
 & Myr
 & $M_{\odot}$
 & Myr
 & $M_{\odot}$
 &
 &
 & \\
\hline
\hline
4 & 96.24 & 0.59 & 28.16 & 0.87 & 0.497 & 0.503 & no \\
5 & 62.78 & 0.81 & 14.84 & 0.91 & 0.507 & 0.493 & no \\
6 & 45.31 & 1.05 & 9.57 & 0.97 & 0.474 & 0.526 & no \\
7 & 34.64 & 1.32 & 6.35 & 1.10 & 0.467 & 0.533 & yes \\
8 & 27.71 & 1.60 & 4.56 & 1.18 & 0.490 & 0.510 & yes \\
9 & 22.87 & 1.88 & 3.53 & 1.33 & 0.497 & 0.503 & yes \\
\hline
\end{tabular}
\tablefoot{
Hydrogen and helium-burning lifetimes and resulting core sizes (of H-exhausted and He-exhausted cores respectively) for our models. The central carbon and oxygen abundances, after core helium-burning are also given. The last columns indicates whether the models ignite carbon. See text for details.
}
\label{tab:main}
\end{center}
\end{table*}


\section{A brief description of MONSTAR and the input physics}

For this work we use MONSTAR, the Monash University Stellar Structure code
(see eg. \cite{fro96,cam08}, and references therein). The code allows one
to compute the evolution of low and intermediate-mass stars from the
pre-main sequence until the last stages of the TP-(S)AGB phase. It has been
used to compute stellar evolution models of metallicity ranging from solar and
 Magellanic Cloud values \citep{kar07, doh10}, to primordial (Z $=0$) and Ultra-Metal-Poor
($\rm{[Fe/H]<-4}$), as defined by \citet{chr05}, and EMP \citep{cam08}.

We treat convective boundaries using the Schwarzschild criterion in
combination with the search for convective neutrality method,
as in \citet{woo81} and \citet{fro96}.  The computation of the dredge-out
phenomenon requires the use of time-dependent mixing and, for the stars
that undergo this process, we used the diffusion equation to
approximate this mixing, as described in \citet{cam08}.

MONSTAR considers only the isotopes that are relevant for the structural
evolution (H, $\rm{^3He}$, $\rm{^4He}$, $\rm{^{12}C}$, $\rm{^{14}N}$,
$\rm{^{16}O}$, and all other species are included in ${Z_{other}}$). The
nuclear reaction rates are from \citet{cau88}.
The ${\rm ^{14}N(p,\gamma)^{15}O}$ rate is updated from the REACLIB value
\citep{run05}.  The computational aspects of the carbon--burning process
are explained in \citet{doh10}. The stellar evolution results from MONSTAR
are usually fed into a post-processing nucleosynthetic code, MONSOON, based
on \citet{can93} and adapted and extended at Monash University
\citep{lat96,lug04}.  The current models experience a large number of
thermal pulses (from a few hundred to more than a thousand). The number of
time steps required to complete the evolution (around 10 million) and, to a
lesser extent, the very high temporal and spatial resolution required,
makes the task of computing the full nucleosynthesis for these models
extraordinarily demanding.  Therefore in this paper we report only the
structural evolution and the yields of some key elements and leave a
further nucleosynthetic study for a future work.  The initial composition
for our model stars is $\rm{X(H)}=0.752$, $\rm{X(He)}=0.248$,
$\rm{X(^{12}C)}=1.8\times 10^{-6}$, $\rm{X(^{14}N})=5.6\times 10^{-7}$ and
$\rm{X(^{16}O)}=5.1\times 10^{-6}$, that is, the scaled-solar composition
from \citet{gre93}, corresponding to {\rm [Fe/H] = -3.2}.

   \begin{figure}
   \centering
   \includegraphics[scale=0.60,angle=0]{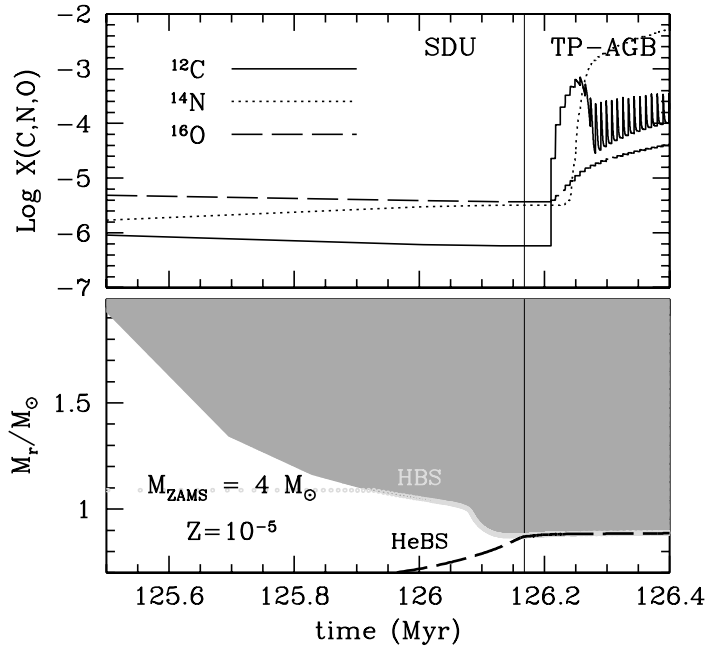}
   \caption{Upper panel: evolution of surface abundances of carbon, nitrogen
and oxygen during the Second Dredge-Up episode (SDU) and the first thermal pulses of the TP-AGB phase
for the 4 ${M_{\odot}}$ model. Lower panel: evolution of the
advance of the base of the convective envelope (grey shading) during the same interval of time.
The light grey and the black dashed lines represent the location of the H-burning shell (HBS) and the 
He-burning shell (HeBS) respectively.
The vertical line marks the beginning of the TP-AGB phase at the onset of the first thermal pulse.}
              \label{fig:surfsdu}
    \end{figure}

   \begin{figure}
   \centering
   \includegraphics[scale=0.60,angle=0]{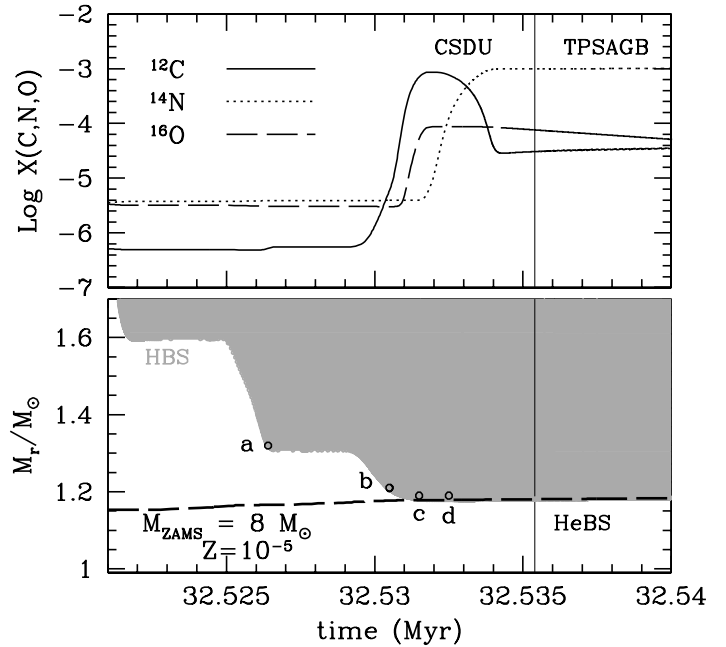}
   \caption{Upper panel: evolution of surface abundances of carbon, nitrogen
and oxygen during the Corrosive Second Dredge-Up episode and the first thermal pulses of the TP-SAGB phase
 for the 8 ${ M_{\odot}}$ model. Lower panel: evolution of the
advance of the base of the convective envelope (grey shading) during the same interval of time.
The lines have the same meaning as in Figure \ref{fig:surfsdu}.
Letters a, b, c and d refer to the times for the corresponding panels of Figure \ref{fig:cartoon_csdu}.}
              \label{fig:csdu} 
    \end{figure}

   \begin{figure}
   \centering
   \vspace{0.5cm}
   \includegraphics[scale=0.90,angle=0]{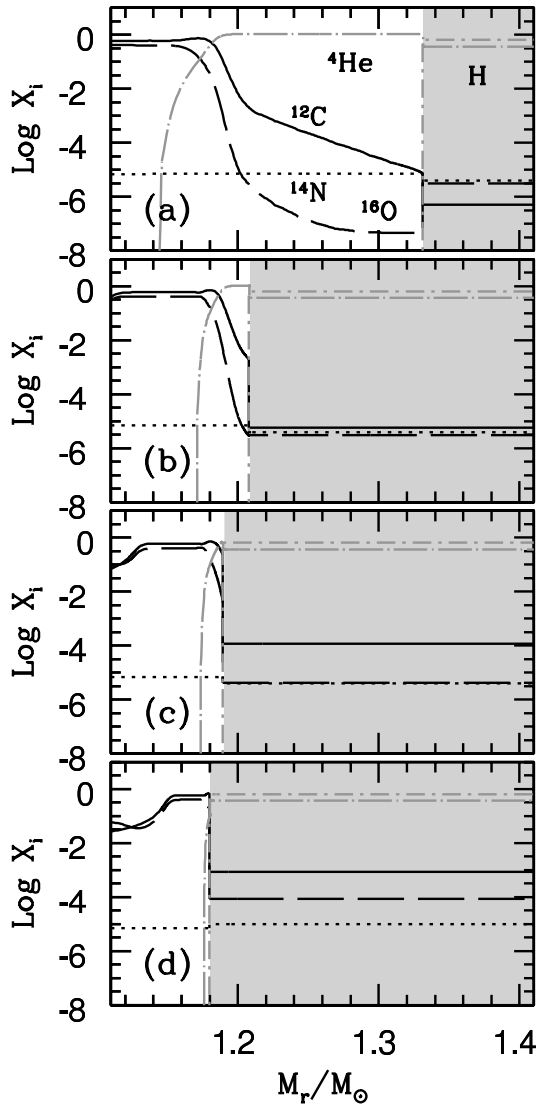}
   \caption{Abundance profiles of carbon, nitrogen
and oxygen showing the progress of the Corrosive Second Dredge-Up episode for the $8 M_{\odot}$ star, at the times labeled in Figure \ref{fig:csdu}. The shaded region represents the convective envelope.}
              \label{fig:cartoon_csdu}
    \end{figure}

MONSTAR uses OPAL opacities \citep{igl96} for the interior, and molecular
opacities that consider the variations of carbon and nitrogen
abundances at the surface \citep{lea09,mar09}. One may expect this
last update to be relevant for the computation of $Z=10^{-5}$ TP-(S)AGB
models.  The reason is that stars of this metallicity experience a very
deep second dredge-up during the early (S)AGB phase that increases
the abundance of carbon at the surface drastically. It has been shown that
these characteristic surface abundances in higher metallicity stars cause a
decrease in surface temperature, an increase in the upper envelope opacity
and, consequently, an increase in surface radius and luminosity; see eg.
\citet{mar02}, \citet{cri07}, \citet{ven10}, \citet{ven12}.  Under these
circumstances the mass loss rates increase, the duration of the TP-(S)AGB
is shortened, and the total yields vary. 

We used the prescription of \citet{vwo93} to compute the mass loss
rates during the early and TP-(S)AGB phase.  It must be noted at this point
that there are many uncertainties about these rates because there are no
observational data to compare to at the metallicity we are considering. It
is usually understood that the most metal-poor stars 
(from primordial metallicity up to Z such that the envelope CNO catalysts after the SDU
are too low to allow efficient H-shell burning) tend to be
more compact and less luminous during the AGB \citep{fuj84}, 
thus their mass loss rates are lower. 
In general for low metallicity stars of different initial masses and 
evolutionary stages, some authors simply apply standard mass-loss rate formulae
and assume these laws also apply at this metallicity \citep{kar07,sie10}. Other authors 
use the so-called Z-scaling, that is,
they multiply the results given by the standard wind prescriptions by a
certain power of the metallicity Z, usually $(Z/Z_{\odot})^{1/2}$
\citep{eld06}, or $(Z/Z_{\odot})^{3/2}$ \citep{vin01}.  We actually find,
as we explain in the next section, that the second dredge-up episode
in intermediate-mass $Z=10^{-5}$ stars leads to surface CNO abundances 
near-solar values and that stellar radii are around 800 $R_{\odot}$,
also similar to those of solar metallicity SAGB stars. Thus we use the same
prescriptions that are applied to solar metallicity stars, with no
modifications.

   \begin{figure}
   \centering
   \vspace{0.3cm}
   \includegraphics[scale=0.60,angle=0]{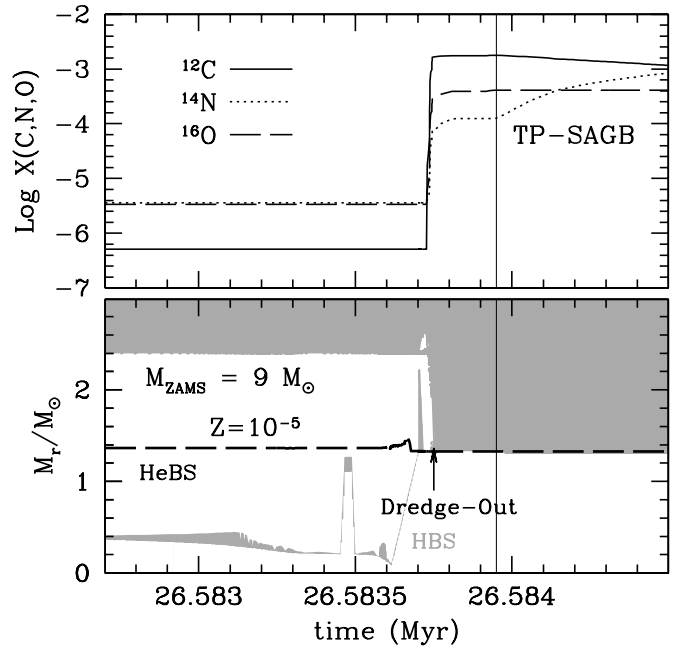}
   \caption{Upper panel: evolution of surface abundances of carbon, nitrogen
and oxygen during carbon burning and the Dredge-Out episode (DO). Lower panel: evolution of the
advance of the base of the convective envelope during carbon-burning and the DO episode
Both panels are for the 9 ${ M_{\odot}}$ star.
The lines have the same meaning as in Figure \ref{fig:surfsdu}.
}
              \label{fig:surfdo}
    \end{figure}

%

\section{Evolution before the TP-(S)AGB phase}

The early evolution of intermediate-mass models with metallicity ${Z
  \la {\rm 10^{-3}}}$ is characterised by both core H- and He-burning
occurring in the blue part of the HR diagram \citep{gir96,chi01}. We can
see this feature in Figure \ref{fig:hrs}, which shows helium ignition
occurs in the blue, soon after hydrogen exhaustion. The loop in the HR
diagram is caused by this ignition of He and it interrupts the usual
evolution to the red giant branch.

In Table \ref{tab:main} we report the initial masses of the computed
$Z = \rm{10^{-5}}$ stars, the core H- and core He-burning lifetimes
(${t_{CHB}}$ and ${t_{CHeB}}$, respectively), the masses of the
hydrogen-exhausted cores after core H-burning (${M_{He}}$), the masses
of the helium-exhausted cores after core He-burning (${M_{CO}}$), and
the central carbon and oxygen abundances at the end of CHeB. We also
indicate which models ignite central carbon.

Table \ref{tab:surf_phases} presents the surface abundances of important
species for our models at different times.  Specifically we give the C, N
and O mass fractions after the end of the SDU or dredge-out (DO), whichever
occurs for the given mass, followed by the (number) ratios C/O and N/O. We
then give the same information at the time when carbon surface abundance
reaches its maximum, which occurs during early phases of the TP-SAGB phase
for models of initial mass $\la$ 6 ${M_{\odot}}$ and during the
SDU or DO for the higher mass models.  Finally we give the same values
after the ${\rm 20^{th}}$ thermal pulse (TP) and for the last calculated model.
The ${\rm 20^{th}}$ pulse is a representative early pulse. For all our models the ${\rm 20^{th}}$ 
pulse occurs after the occurrence of the local maximum in carbon surface abundance
and its corresponding abundances (by comparison with those at the end of the SDU/DO)
reflect that both the TDU and HBB are operative from the first pulses.
The high abundances seen in the 8 and 9 ${M_{\odot}}$ are a consequence
of the corrosive second dredge-up episode that we explain in the next
subsection. Such surface CNO enrichment in primordial stars has been
reported by other authors, e.g. \citet{chi01}, \citet{sie02},
\citet{gil07}, \citet{lau09} and \citet{sud10}.

   \begin{figure}
   \centering
   \hspace{-0.2cm}
   \includegraphics[scale=0.60,angle=0]{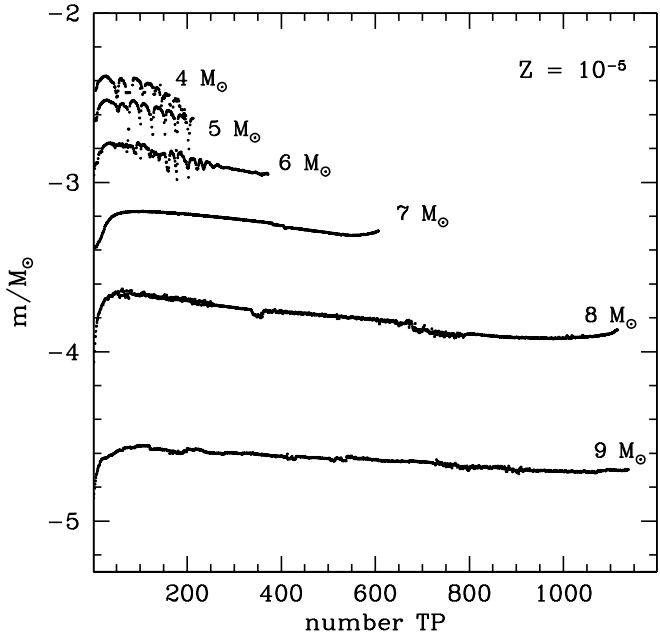}
   \caption{Mass of intershell convective zones versus pulse number for our models.}
              \label{fig:ishell}
    \end{figure}

\subsection{The Corrosive Second Dredge-Up Episode}
Once helium is exhausted in the core, the model stars begin ascending the
second\footnote{This is actually the first and only ascension of the giant
  branch for these low-metallcity stars since they avoid the RGB. We retain the
  terminology for continuity with higher metallicity stars.} giant branch
-- the AGB. At this phase the models undergo the second dredge-up episode
(SDU)\footnote{Similarly this is actually the first dredge-up episode for
  low-metallicity stars but again we retain the terminology.}.  

Models with
core masses $\la 1.1$ M$_{\odot}$ after the SDU are unable to burn
carbon and evolve along the TP-AGB phase \citep{nom84,nom87,sie07}.  Models
with core masses $\gtrsim 1.1$ M$_{\odot}$ after the SDU go on to ignite
carbon and, as we see in the following sections, either enter the
TP-SAGB phase to become white dwarfs or electron-capture SNe (hereafter
EC-SNe), or evolve directly to core-collapse supernovae.  The lower initial
mass limit for C-burning at $Z = {\rm 10^{-5}}$ is about 7.0 $M_{\odot}$, but
this value changes for different metallicities and input physics
\citep{gil03,sie07,gil10}.

For those models that ignite carbon, the SDU episode occurs either before
(for initial masses up to 7.5-8 M$_{\odot}$) or after carbon burning (for
initial masses above 8 M$_{\odot}$). Standard SDU is characterised by the advance
inwards of the convective envelope towards regions that have been processed
by the CN(O) cycle, and therefore it manifests itself as an increase in N
surface abundances and a decrease in C and O surface abundances (see Figure
\ref{fig:surfsdu}).

The SDU has an interesting peculiarity in the case of EMP stars of masses
$\ga 8 {M_{\odot}}$ (see Figure \ref{fig:csdu}).  In these stars
the convective envelope reaches beyond regions processed by
H burning and it is able to penetrate to regions where He burning has
been active, as seen in Figure \ref{fig:cartoon_csdu}. Therefore the SDU is
 able to transport not only the products of CNO nucleosynthesis, but also the
products of He burning. This was also reported by \citet{sud10}. Due to this phenomenon 
the effective Z (dominated now by C+N+O)
increases by several orders of magnitude (see Table
\ref{tab:surf_phases}). We call this process the
Corrosive SDU episode (Doherty et al. 2013, in preparation).  It is
also interesting to see in Figure \ref{fig:csdu} that, just after the
Corrosive SDU but before the TP-SAGB begins, the surface abundance of
nitrogen increases, while the carbon abundance and, to a lesser extent,
oxygen abundance, decrease. This is caused by 
the early onset of hot-bottom-burning (HBB).


\begin{table*}
\caption{
CNO surface composition values and quotients at selected times along the evolution of our model stars.
}
\begin{center}
\begin{tabular}{ccccc}
&
&
{At end of SDU/DO}&
&
\\
\end{tabular}

\begin{tabular}{ccccccccc}
\hline
\hline
{${M_{ZAMS}/M_{\odot}}$}
  & {${X(C)}$}
  & {${X(N)}$}
  & {${X(O)}$}
  & {$ X(C+N+O)$}
  & {${C/O}$}
  & {${N/O}$}\\
\hline
4.0  & $5.8\times 10^{-7}$ & $3.1\times 10^{-6}$ & $3.9\times 10^{-6}$  & $7.6\times 10^{-6}$ & 0.20 & 0.91  \\
5.0  & $5.3\times 10^{-7}$ & $3.5\times 10^{-6}$ & $3.4\times 10^{-6}$  & $7.4\times 10^{-6}$ & 0.20 & 1.18  \\
6.0  & $4.9\times 10^{-7}$ & $3.8\times 10^{-6}$ & $3.1\times 10^{-6}$  & $7.4\times 10^{-6}$ & 0.21 & 1.40  \\
7.0  & $4.1\times 10^{-6}$ & $4.0\times 10^{-6}$ & $3.1\times 10^{-6}$  & $1.1\times 10^{-5}$ & 1.72 & 1.72  \\
8.0  & $2.7\times 10^{-5}$ & $9.0\times 10^{-4}$ & $8.2\times 10^{-5}$  & $1.0\times 10^{-3}$ & 0.44 & 12.54 \\
9.0  & $1.6\times 10^{-3}$ & $6.5\times 10^{-5}$ & $2.8\times 10^{-4}$  & $2.0\times 10^{-3}$ & 5.71 & 0.23  \\
\hline
\\
\end{tabular}

\begin{tabular}{ccccc}
&
&
{At maximum ${X_{surf}(C)}$}
&
\\
\end{tabular}

\begin{tabular}{ccccccccc}
\hline
\hline
{${M_{ZAMS}/M_{\odot}}$}
  & {stage}
  & {${X(C)}$}
  & {${X(N)}$}
  & {${X(O)}$}
  & {$ X(C+N+O)$}
  & {${C/O}$}
  & {${N/O}$}\\
\hline
4.0 & ${\rm 12^{th}}$ TP & $6.7\times 10^{-4}$ & $1.8\times 10^{-4}$ & $1.3\times 10^{-5}$ & $8.6\times 10^{-4}$ & 69.0 & 15.8  \\
5.0 & ${\rm 13^{th}}$ TP & $2.5\times 10^{-4}$ & $8.3\times 10^{-5}$ & $6.9\times 10^{-6}$ & $3.4\times 10^{-4}$ & 48.0 & 13.7  \\
6.0 & ${\rm 13^{th}}$ TP & $3.9\times 10^{-5}$ & $7.9\times 10^{-6}$ & $3.6\times 10^{-6}$ & $5.1\times 10^{-5}$ & 14.4 & 2.51  \\
7.0 & SDU & $4.2\times 10^{-6}$ & $3.9\times 10^{-6}$ & $3.0\times 10^{-6}$ & $1.1\times 10^{-5}$ & 1.72 & 1.72  \\
8.0 & SDU & $8.5\times 10^{-4}$ & $4.3\times 10^{-6}$ & $8.0\times 10^{-5}$ & $9.3\times 10^{-4}$ & 26.4 & 0.06  \\
9.0 & DO & $1.7\times 10^{-3}$ & $1.2\times 10^{-4}$ & $4.0\times 10^{-4}$ & $2.2\times 10^{-3}$ & 4.25 & 0.30 \\
\hline
\\
\end{tabular}

\begin{tabular}{ccccc}
&
&
{After 20 TPs}
&
\\
\end{tabular}

\begin{tabular}{ccccccccc}
\hline
\hline
{${M_{ZAMS}/M_{\odot}}$}
  & {${X(C)}$}
  & {${X(N)}$}
  & {${X(O)}$}
  & {$ X(C+N+O)$}
  & {${C/O}$}
  & {${N/O}$}\\
\hline
4.0 & $3.2\times 10^{-4}$ & $2.7\times 10^{-3}$ & $2.7\times 10^{-5}$ & $3.0\times 10^{-3}$ & 15.7 & 111.8 \\
5.0 & $1.5\times 10^{-4}$ & $1.2\times 10^{-3}$ & $1.4\times 10^{-5}$ & $1.4\times 10^{-3}$ & 13.9 &  99.0 \\
6.0 & $4.9\times 10^{-5}$ & $3.7\times 10^{-4}$ & $4.3\times 10^{-6}$ & $4.2\times 10^{-4}$ & 15.0 &  96.9 \\
7.0 & $3.8\times 10^{-7}$ & $1.1\times 10^{-5}$ & $6.0\times 10^{-8}$ & $1.2\times 10^{-5}$ & 8.4  &   7.2 \\
8.0 & $3.4\times 10^{-5}$ & $1.0\times 10^{-3}$ & $5.3\times 10^{-5}$ & $1.1\times 10^{-3}$ & 0.9  &  21.7 \\
9.0 & $5.3\times 10^{-4}$ & $1.6\times 10^{-3}$ & $4.1\times 10^{-4}$ & $2.5\times 10^{-3}$ & 1.7  &   4.4 \\
\hline
\\
\end{tabular}

\begin{tabular}{ccccc}
&
&
{At end of calculations}
&
\\
\end{tabular}

\begin{tabular}{ccccccccc}
\hline
\hline
{${M_{ZAMS}/M_{\odot}}$}
  & {${X(C)}$}
  & {${X(N)}$}
  & {${X(O)}$}
  & {$ X(C+N+O)$}
  & {${C/O}$}
  & {${N/O}$}\\
\hline
4.0 & $2.3\times 10^{-3}$ & $4.4\times 10^{-2}$ & $4.0\times 10^{-4}$ & $4.7\times 10^{-2}$ &  7.7 & 125.7 \\
5.0 & $2.8\times 10^{-3}$ & $4.1\times 10^{-2}$ & $3.8\times 10^{-4}$ & $4.4\times 10^{-2}$ & 9.6 & 123.3  \\
6.0 & $2.0\times 10^{-3}$ & $3.0\times 10^{-2}$ & $2.7\times 10^{-4}$ & $3.2\times 10^{-2}$ & 10.2 & 131.0 \\
7.0 & $1.3\times 10^{-3}$ & $1.7\times 10^{-2}$ & $1.9\times 10^{-4}$ & $1.9\times 10^{-2}$ &  9.4 & 100.0 \\
8.0 & $7.4\times 10^{-4}$ & $6.3\times 10^{-3}$ & $7.5\times 10^{-5}$ & $7.1\times 10^{-3}$ & 13.2 &  96.0 \\
9.0 & $1.2\times 10^{-4}$ & $2.3\times 10^{-3}$ & $9.5\times 10^{-5}$ & $2.5\times 10^{-3}$ & 1.7 &  27.7 \\
\hline
\end{tabular}
\tablefoot{ 
Values are given at the end of the second dredge-up or dredge-out episode, 
at the time when surface carbon abundance reaches its maximum value,
at the $\rm{20^{th}}$ thermal pulse of the TP-(S)AGB phase 
and at the end of our calculations. For the maximum ${X_{surf}(C)}$ entries the column labelled "stage" tells when this maximum is reached.}
\label{tab:surf_phases}
\end{center}
\end{table*}


\subsection{The Dredge-Out Episode}

It is well known that at the upper mass range of SAGB stars there is a
complex interplay of overlapping convective zones.  It has been common to
refer to ``dredge-out'' in the case where the outer convective region (rich
in hydrogen) makes contact with the inner He-burning convective region at
the end of carbon burning \citep{rit99,sie07,gil10}.  This results in a
proton ingestion episode (PIE), as described by eg. \citet{cam08} in the
context of lower mass stars.

The high CNO surface abundances seen in our 9 ${M_{\odot}}$ model are a
result of such behaviour, as shown in Figure 5.  In this model we see
convection due to C burning in a shell in the inner part of the star (M
$\sim$ 0.4 ${M_{\odot}}$). Pockets of convective He burning arise,
notably at time $\sim$ 26.5837 Myr and mass $\sim$ 1.4-2 ${M_{\odot}}$.
Immediately after this there is a rapid advance of the hydrogen-rich
envelope downward.  Crucially this convective zone joins with the He
burning convective region below, causing a rapid ingestion of protons into
regions of very high temperature so that a hydrogen flash (with ${L_H}$
as high as $10^{10}{L_{\odot}}$) occurs.  In time scales of months
intense H- and He-burning occurs at very high rates (C burning continues at
inner regions), and the products of the CNO-cycle and the 3$\alpha$
reactions are mixed to the surface.

\section{Evolution during the TP-(S)AGB phase}

Following the SDU or DO episodes, the stars proceed to the TP-(S)AGB.  
From Table 2 we see that either due to corrosive SDU or to the DO process,
the envelopes of our 8 and 9 ${M_{\odot}}$ stars are sufficiently
enriched in CNO isotopes that their total surface metallicities at the
beginning of the TP-SAGB phase are all higher than $10^{-3}$. Stars of 4
and 5 ${M_{\odot}}$ experience early and very deep Third Dredge-Up
(TDU), combined with strong hot bottom burning, which leads to a similar
result.  After the first 10 thermal pulses the surface abundances of
carbon, ${X_{surf}(C)}$, are higher than $10^{-4}$, and keep increasing
until they reach a local maximum at the ${\rm 12^{th}}$ and the ${\rm
  13^{th}}$ pulses for the 4 and 5 ${M_{\odot}}$ models
respectively. HBB is responsible for an increase in surface nitrogen
abundance to values ${X_{surf}(N) > 10^{-3}}$ at their ${\rm 20^{th}}$
thermal pulses, as shown in Table 2.
The 6 and 7 ${M_{\odot}}$ models experience more moderate SDU and early
TDU episodes.  Thus their total surface metallicities at their ${\rm
  20^{th}}$ pulses are comparatively low.  Nevertheless the combination of
TDU and HBB eventually increases their ${X_{surf}(N)}$ to values higher
than ${\rm 10^{-3}}$ at the ${\rm 30^{th}}$ and the ${\rm 60^{th}}$ pulses,
respectively.  Considering that the corresponding models undergo 372 and
607 thermal pulses, we can say that these stars also host envelopes
enriched in C, N and O during most of their TP-(S)AGB.  Therefore the
hydrogen-burning shells (HBS) in our models burn efficiently via
CNO-cycling in a very similar way to that of higher metallicity stars and
the structural characteristics of the envelopes are also very similar to
those of stars of higher metallicity.

Our calculations halt when an envelope instability of the type studied by
\citet{lau12}, \citet{pet06} or \citet{wag94} is encountered. The code can
not converge because of the lack of a hydrostatic solution which occurs due
to the region just above the hydrogen shell exceeding the Eddington
luminosity.  When this occurs the models still have envelopes ranging from
0.25 ${M_{\odot}}$ for the 4 ${M_{\odot}}$ model, and up to 1.5
${M_{\odot}}$ for the 9 ${M_{\odot}}$ model. As a result of this
instability some considerable uncertainty is introduced into the
predictions for the final fates of the stars as well as the yields, as we
discuss in the following sections.

   \begin{figure*}
   \centering
   \includegraphics[scale=0.65,angle=0]{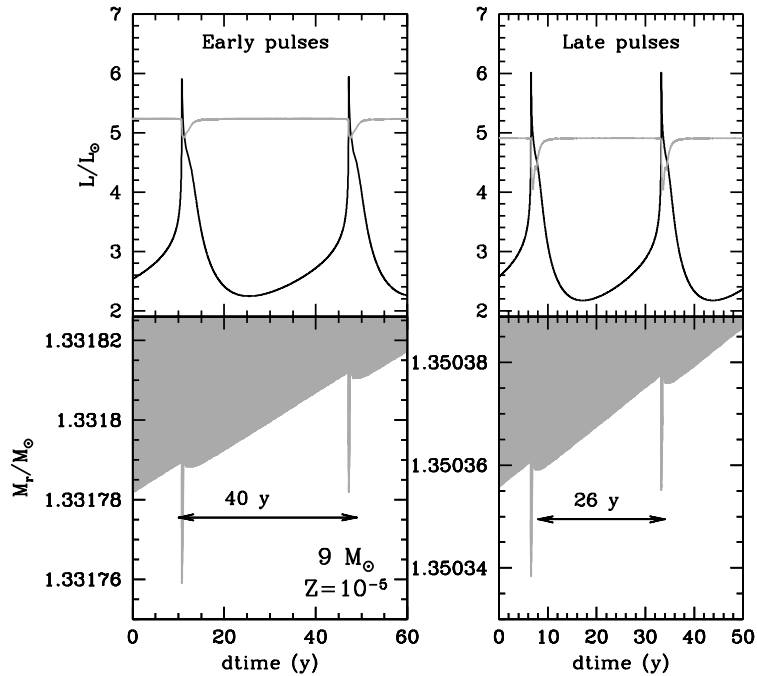}
   \caption{Upper left panel: Evolution of the luminosity associated with hydrogen and helium burning for the $9 M_{\odot}$ model star, during pulses 125 and 126.
Lower left panel: Evolution of convective regions during the same pulses. Right panels represent the same as left panels for the case of pulses 1125 and 1126.}
    \label{fig:tpsagb}
    \end{figure*}

These massive (S)AGB stars begin their thermally pulsing evolution with relatively
large cores, and hence small interpulse periods, which decrease further
during the evolution. For the 4 ${M_{\odot}}$ model the interpulse
decreases from about 10000 years to 3000 years in 200 thermal pulses, while
the 8 and 9 ${M_{\odot}}$ models have, respectively, periods of only
about 200 years and 40 years between pulses, which number over 1000 before
the end of the calculations.  Similarly, the intershell convective region
is exceptionally thin in SAGB stars, extending only about 0.0002 ${
  M_{\odot}}$ in the 8 ${ M_{\odot}}$ model, for example, although it
increases to about 0.004 for masses as low as 4 $ M_{\odot}$, as shown
in Figure \ref{fig:ishell}.

The temperature in the intershell convective region is very high, being
about 360 million degrees for all masses. This is sufficiently high for a
strong activation of the ${\rm ^{22}Ne}$ neutron source and hence
substantial s-processing is expected. The details of this neutron capture
nucleosynthesis will be the subject of a later paper.  For now we note that
the thin intershell means that only small amounts of material are dredged
up to the surface following each pulse, even though the dredge-up parameter
$\lambda$ may be quite large (near 1 for initial masses ${M} \la
6$ ${ M_{\odot}}$, decreasing to about 0.5 for the 8 $ M_{\odot}$
model and 0.05 for 9 $ M_{\odot}$ model). The surface of the star will
be nevertheless enriched in s-process elements, which is expected to be
noticeable because of the very small initial reservoir of such species
present in models with such low metallicity.

\begin{table*}[t]
\caption{
Relevant results from the evolutionary calculations that characterise the TP-(S)AGB of our models.
}
\begin{center}
\begin{tabular}{cccccccccccc}
\hline
{${M_{ZAMS}}$}
  & {Number TP}
  & {${<M_{csh}>}$}
  & {${<M_{du}>}$}
  & {${<\lambda>}$}
  & {${<T_{HeBS}>}$}
  & {${<T_{HBS}>}$}
  & {${<T_{BCE}>}$}
  & {${<\Delta t_{IP}>}$}
  & {$ M_{core,f}$}
  & {$ M_{env,f}$}\\
{${M_{\odot}}$}
  & ~
  & {$M_{\odot}$}
  & {$M_{\odot}$}
  & ~
  & {MK}
  & {MK}
  & {MK}
  & {yr}
  & {$M_{\odot}$}
  & {$M_{\odot}$}\\
\hline
\hline
4.0 &  197 &  $3.6\times 10^{-3}$ &  $2.7\times 10^{-3}$ &  0.92 &  359.1 &  90.5 &  84.7 &  6702 & 0.93 & 0.25 \\
5.0 &  213 &  $2.6\times 10^{-3}$ &  $1.9\times 10^{-3}$ &  0.91 &  356.7 &  95.1 &  88.4 &  4382 & 0.97 & 0.54 \\
6.0 &  372 &  $1.4\times 10^{-3}$ &  $9.8\times 10^{-4}$ &  0.85 &  361.8 & 104.1 &  98.7 &  1828 & 1.04 & 0.69 \\
7.0 &  607 &  $5.8\times 10^{-4}$ &  $3.9\times 10^{-4}$ &  0.78 &  366.8 & 116.5 & 113.6 &   739 & 1.14 & 0.73 \\
8.0 & 1114 &  $1.6\times 10^{-4}$ &  $6.2\times 10^{-5}$ &  0.48 &  363.1 & 130.8 & 127.6 &   183 & 1.26 & 0.53 \\
9.0 & 1138 &  $2.3\times 10^{-5}$ &  $1.1\times 10^{-5}$ &  0.05 &  362.4 & 147.5 & 144.7 &    30 & 1.35 & 1.52 \\
\hline
\end{tabular}
\tablefoot{
   Angled brackets indicate linear averages over the total
   number of pulses. From left to right we give: the initial mass,
   the number of computed thermal pulses, the mass of the intershell
   convective region, the amount of material dredged-up after each
   pulse, the dredge-up parameter, the maximum temperature reached in
   the inner convective region at each pulse, the maximum temperatures at the base 
   of the H-shell (defined as the point at which H-mass fractions $\la$ 0.05) 
   and at the base of the convective envelope at each interpulse, 
   the interpulse period (defined as the time between two consecutive pulses p+1 and p) 
   and the final core and envelope masses at the end of our calculations.
   }
\label{tab:lambda}
\end{center}
\end{table*}

The evolution during the TP-SAGB phase of the 9 $M_{\odot}$ model is
shown in Figure \ref{fig:tpsagb}.  On the left we show two consecutive
early pulses, numbers 125 and 126, while the right hand panel shows two
consecutive pulses much later in the evolution, being thermal pulse 
numbers 1125 and 1126.  We can see the expected increase in the peak luminosity of the He
shell in later pulses, but we also note a new behaviour. The hydrogen
burning luminosity, which in low mass stars drops to essentially zero
during dredge-up, here remains reasonably high. 
It does decrease with later
pulses, as seen in the diagram, but surprisingly the shell is not
extinguished during the brief dredge-up phase itself. This plays a role in
terminating the dredge-up earlier than would be expected normally for
massive (S)AGB stars, keeping $\lambda$ as low as 0.05 compared to values near
unity for stars at the lower end of our mass range (eg 4 $ M_{\odot}$).
A similar behaviour has been described by \citet{chi01} and by \citet{her04a}, who named it Hot-TDU.
We note in passing that Figure \ref{fig:tpsagb} highlights the need for
high resolution in both space and time.  The duration of each convective
shell associated with helium flashes is as short as a few months and its
extent is typically a few $10^{-5} M_{\odot}$.

The main nucleosynthesis process acting in these stars is hot bottom
burning with temperatures at the base of the convective envelope above 140
million degrees. These temperatures are enough for efficient CNO cycling as
well as the activation of the Ne-Na and Mg-Al reactions.

\begin{figure}
   \centering
   \includegraphics[scale=0.60,angle=0]{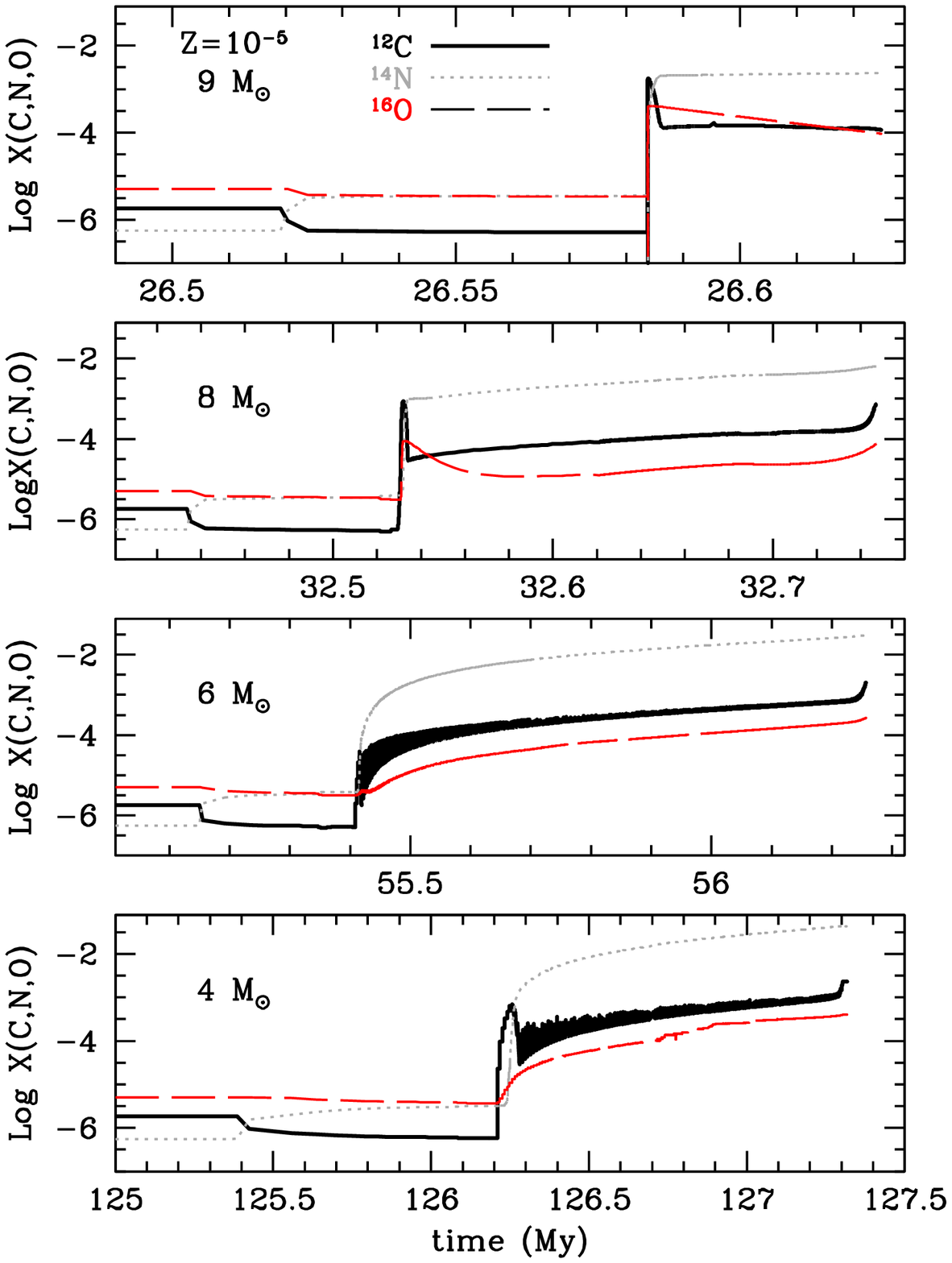}
   \caption{Evolution of surface abundances
   versus time for the 4, 6, 8 and 9 $M_{\odot}$ models.
   }
              \label{fig:evosurf}
    \end{figure}

The evolution of surface abundances of carbon, nitrogen and oxygen along
the TP-(S)AGB for the 4, 6, 8 and 9 $ M_{\odot}$ models is show in
Figure \ref{fig:evosurf}. We note that, either due to the corrosive SDU 
or to the TDU, the number ratio C/O remains higher
than one during most of the TP-(S)AGB phase of the stars considered,
despite the very strong HBB.  Table
\ref{tab:lambda} shows some relevant results from the evolutionary
calculations along the TP-(S)AGB of our models.

  \begin{figure}
   \centering
   \vspace{-0.3cm}
   \includegraphics[scale=0.60,angle=0]{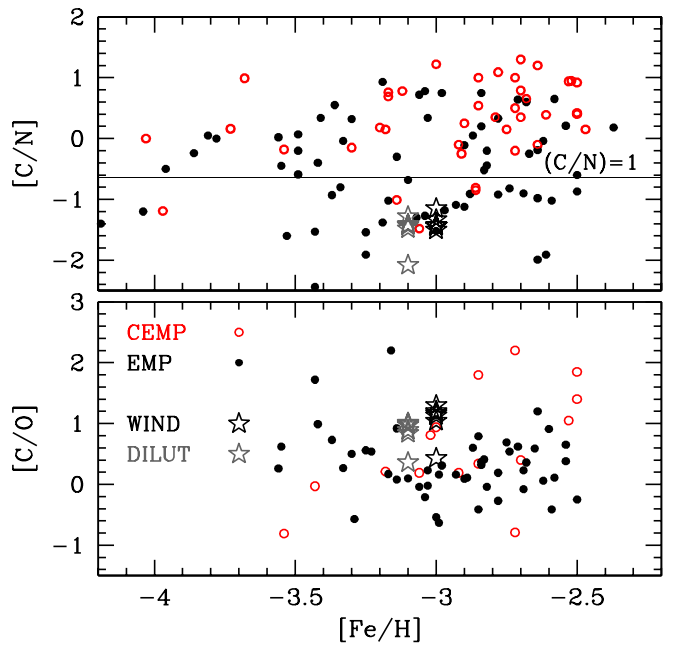}
   \caption{${\rm [C/N]}$ (upper panel) and  ${\rm [C/O]}$ (lower panel) versus metallicity.
   Black symbols correspond to the yields of our computed models.
The cases in which dilution has been taken into account are represented in grey. They should be 
located at the same [Fe/H] as the non-diluted yields, but we displaced them -0.1 dex for the sake of clarity.
   For comparison, observational data of EMP (black solid circles) and CEMP (red open circles) from the SAGA database are
   represented in both panels.}
              \label{fig:cno}
    \end{figure}

 \begin{figure}[t]
   \centering
   \vspace{-0.1cm}
   \includegraphics[scale=0.60,angle=0]{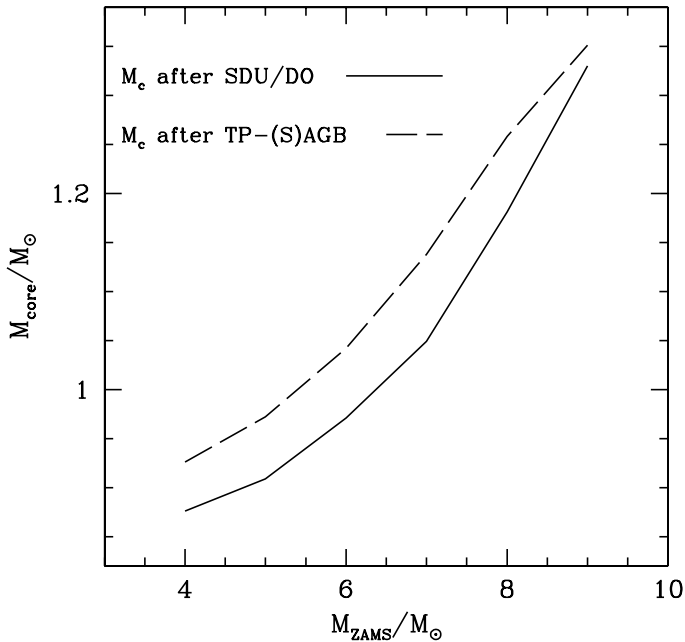}
   \caption{ Masses of degenerate cores versus initial mass, at the end of the second dredge-up or dredge-out episode and at the end of our full computations for our model stars.}
              \label{fig:mcores}
    \end{figure}

\section{Yields and final fates of single intermediate-mass ${\bf Z=10^{-5}}$ stars.}

\begin{table*}
\caption{Total ejected envelope masses and hydrogen, helium, carbon, nitrogen and oxygen yields from our model stars.}
\begin{center}
\begin{tabular}{crrrrrrrr}
\hline
  {${ M_{ZAMS}/M_{\odot}}$}
  & { 4 ${ M_{\odot}}$}
  & { 5 ${ M_{\odot}}$}
  & { 6 ${ M_{\odot}}$}
  & { 7 ${ M_{\odot}}$}
  & { 8 ${ M_{\odot}}$}
  & { 9 ${ M_{\odot}}$} 
  & {$\Delta$ Yields}  \\
\hline
\hline
\vspace{0.1cm}
${ M_{env}}$ & $3.08$ & $3.93$ & $4.96$ & $5.85$ & $6.74$ & $7.65$ & - \\
${ M_{ej}(H)}$  & $1.68$ & $2.02$ & $2.61$ & $3.38$ & $4.12$ &  $5.21$ & Lau \\
\vspace{0.1cm}
${ M_{ej}(H)}$  & $1.68$ & $2.02$ & $2.61$ & $3.38$ & $4.12$ &  $5.21$ & Syn \\
${ M_{ej}(He)}$ & $1.27$ & $1.75$ & $2.20$ & $2.48$ & $2.62$ &  $2.43$ & Lau \\
\vspace{0.1cm}
${ M_{ej}(He)}$ & $1.27$ & $1.75$ & $2.20$ & $2.48$ & $2.62$ &  $2.44$ & Syn \\
${ M_{ej}(C)}$ & $4.37\times 10^{-3}$ & $5.60\times 10^{-3}$ & $4.82\times 10^{-3}$ & $3.21\times 10^{-3}$ & $1.39\times 10^{-3}$ & $1.19\times 10^{-3}$ & Lau \\
\vspace{0.1cm}
${ M_{ej}(C)}$ & $4.37\times 10^{-3}$ & $5.60\times 10^{-3}$ & $5.40\times 10^{-3}$ & $4.25\times 10^{-3}$ & $1.79\times 10^{-3}$ & $1.10\times 10^{-3}$ & Syn \\

${ M_{ej}(N)}$ & $1.25\times 10^{-1}$ & $1.50\times 10^{-1}$ & $1.41\times 10^{-1}$ & $8.66\times 10^{-2}$ & $3.00\times 10^{-2}$ & $1.69\times 10^{-2}$ & Lau \\
\vspace{0.1cm}
${ M_{ej}(N)}$ & $1.25\times 10^{-1}$ & $1.50\times 10^{-1}$ & $1.41\times 10^{-1}$ & $8.66\times 10^{-2}$ & $3.00\times 10^{-2}$ & $1.99\times 10^{-2}$ & Syn \\

${ M_{ej}(O)}$ & $1.13\times 10^{-3}$ & $1.30\times 10^{-3}$ & $1.04\times 10^{-3}$ & $7.41\times 10^{-4}$ & $2.22\times 10^{-4}$ & $1.37\times 10^{-3}$ & Lau \\
\vspace{0.1cm}
${ M_{ej}(O)}$ & $1.13\times 10^{-3}$ & $1.30\times 10^{-3}$ & $1.05\times 10^{-3}$ & $7.67\times 10^{-4}$ & $2.22\times 10^{-4}$ & $9.71\times 10^{-4}$ & Syn \\

\hline
\hline
${\rm [C/Fe]}$ & 2.9 & 2.9 & 2.8 & 2.5 & 2.1 & 2.0 & Lau \\
${\rm [N/Fe]}$ & 4.8 & 4.8 & 4.7 & 4.4 & 3.9 & 3.6 & Lau \\
${\rm [O/Fe]}$ & 1.9 & 1.8 & 1.6 & 1.4 & 0.8 & 1.5 & Lau \\
\hline
${\rm [C/Fe]_{dil}}$ & 1.9 & 2.0 & 2.0 & 1.8 & 1.4 & 1.3 & Lau \\
${\rm [N/Fe]_{dil}}$ & 3.8 & 3.9 & 3.9 & 3.6 & 3.2 & 2.9 & Lau \\
${\rm [O/Fe]_{dil}}$ & 0.9 & 0.9 & 0.9 & 0.7 & 0.3 & 0.9 & Lau \\

\hline
\end{tabular}
\tablefoot{
The last column indicates whether the yields were computed under the instant mass loss (Lau) or the
synthetic approximation (Syn) -see full text for details. The last three rows show the effect of
diluting $1\%$ of the ejected matter in a 0.3 ${ M_{\odot}}$ envelope of a 0.8 ${ M_{\odot}}$ secondary star.
}
\label{tab:winds}
\end{center}
\end{table*}

We computed the yields of hydrogen, helium, carbon, nitrogen and
oxygen, according to the expression:
\begin{equation}
{ M_i = \int_0^\tau  X_i(t) \: \dot M_{env}(t) \: dt}
\end{equation}

\noindent where ${ M_i}$ is the ejected mass of element i, ${
  X_i(t)}$ is the mass fraction of that element at a certain time t, ${
  M_{env}(t)}$ is the envelope mass at t and $\tau$ is the time at the end
of our calculations.

Because non-negligible envelope masses remain at the end of calculations,
the total yields we present are computed in two different ways.  First we
assume that, immediately after the \citet{lau12} instability is reached,
the envelope mass is lost on dynamical time scales. Thus the remnant yield
of element i at this time is
\begin{equation}
{ \Delta M_i = X_i(\tau) \: M_{env}(\tau).}
\end{equation}

As a second option we assume that the stars are able to recover after the
\citet{lau12} instability and proceed along further thermal pulses that
mimic the characteristics (duration, average mass loss rates) of the last
computed thermal pulse.  This is the spirit of the synthetic approach used
by \citet{kar07} and references therein, which uses the dredge-up parameter
of the last computed model to calculate the variations of the envelope
composition.  Because in our most massive models hot-bottom burning 
is still moderately active, we considered the combined effect of
composition variations due to TDU and HBB simply by getting the envelope
mass fraction variation of each element between the last two thermal pulses
and then iterating until the envelope mass is zero.  We proceeded as
follows.

The envelope mass at pulse p+1 is computed as
\begin{equation}
{ M_{env}(p+1)=M_{env}(p)+\Delta M_{env}}
\end{equation}
\noindent where
\begin{equation}
{ \Delta M_{env}= \dot M_{wind}\:\Delta t_{IP} - \dot M_{core}\:\Delta t_{IP}.}
\end{equation}
${\dot M_{wind}}$ and ${ \dot M_{core}}$ are the mass loss rate due
to stellar winds and the core growth rate respectively, and ${ \Delta
  t_{IP}}$ is the time between two consecutive thermal pulses. Because
${ \dot M_{wind} \approx -10^{-4} M_{\odot}/yr}$ and ${ \dot M_{core}
  \approx 10^{-7} M_{\odot}/yr}$ in our models, the variation of the
envelope mass is dominated by the effect of the winds.

The yield of element i at pulse p+1 is
\begin{equation}
{ Y_i(p+1)=Y_i(p)+X_i(p) \: \Delta M_{env}.}
\end{equation}

The envelope mass fraction variation is obtained using the masses of
element i at the envelope for two consecutive pulses, ${ M_i(p-1)}$ and
${ M_i(p)}$:
\begin{equation}
{ \Delta X_i(p) = X_i(p)-X_i(p-1).
}
\end{equation}
Thus the envelope mass fraction of i at the thermal pulse p+1 is
\begin{equation}
{ X_i(p+1)=X_i(p)+\Delta X_i(p).}
\end{equation}
The numbers of additional thermal pulses we obtain are: 2 for the 4 ${ M_{\odot}}$,
 6 for the 5 ${ M_{\odot}}$, 13 for the 6 ${ M_{\odot}}$,
21 for the 7 ${ M_{\odot}}$, 46 for the 8 ${ M_{\odot}}$ and 576 for the 9 ${ M_{\odot}}$ model.

Our yield results are presented in Table \ref{tab:winds}.  In the mass
range studied the dominance of nitrogen shows the relevance of hot-bottom
burning during most of the TP-(S)AGB evolution. 
The occurrence of corrosive second dredge-up, dredge-out and third dredge-up 
episodes also cause relatively high carbon yields (higher than the oxygen 
yields for all our models except the 9 ${ M_{\odot}}$ case).
 For most elements the
differences in the final yields using the two approaches described above is
very small, although we can see that the 6, 7 and 8 ${ M_{\odot}}$ cases
present C yields which are ${\rm 11\%}$, ${\rm 24 \%}$ and ${\rm 22\%}$
higher when the synthetic treatment is used.  Also O yields are higher, but
their increase is less important. These higher C and O yields are the
result of the TDU process, which is active during our last computed pulses.
The effect of the synthetic pulse treatment on the 9 ${ M_{\odot}}$
model is different, as TDU is very mild (${ \lambda}$=0.05). Instead,
HBB acting along 562 additional synthetic pulses causes an increase of
${\rm 15\%}$ in N from ON cycling.
It should be recalled that the synthetic approach we have described is a simplification. 
HBB tends to become milder (or extinct) as the envelope mass and the temperature at its base
decrease at the last stages of the TP-(S)AGB phase. TDU, on the other hand, remains active
(and increasingly efficient) at these last stages, so we
might be underestimating the effects of TDU and overestimating
the effects of HBB.

The lower part of Table \ref{tab:winds} presents the same yields (with the assumption 
that the instability ends the evolution) using the notation ${\rm
  [A/B]=Log_{10}(N_A/N_B)_* - Log_{10}(N_A/N_B)_{\odot}}$. In this
expression ${\rm (N_A/N_B)_*}$ refers to our final number fractions and
${\rm (N_A/N_B)_{\odot}}$ refers to solar number fractions.


Most of the matter ejected by our model stars is expected to contribute to
the enrichment of the ISM. The importance of such a contribution depends on
the nature of the primordial Galactic IMF which, as we mentioned in
the Introduction, is still an open question. A fraction of stars in the
mass range studied here would have been primary components of wide binaries
and therefore part of the matter ejected through their stellar winds may
have been accreted by their low mass companions, polluting their
surfaces. This is the standard scenario for the formation of CEMP stars,
proposed by \citet{sud04} and \citet{luc05}, and supported by their
observational data, which are consistent with all CEMP-s stars belonging to
binary systems.

We considered the scenario where our TP-(S)AGB model stars might have
been polluters of currently observed CEMP stars. Given the facts that
observed CEMP stars have masses $<$ 1 ${ M_{\odot}}$ and that we are
considering intermediate-mass stars in the high mass range, the mass
quotient, q, between primary and secondary component masses, would have
been $\geq$ 4. High q values in close binaries that experience mass
transfer due to Roche lobe overflow are associated with orbital
instability, the occurrence of common envelopes and, eventually, with the
merger of the components.  However mergers can be avoided if the
interacting binary is wide enough to avoid Roche lobe overflow. In this
scenario mass transfer via stellar winds could pollute the
companion. Keeping this idea in mind, we compared our yields to
observational data from the SAGA database \citet{sud08}, simply to check if
our results fit within the ranges given by observations.  Our results for
the abundance quotients ${\rm [C/N]}$ and ${\rm [C/O]}$ are presented in
Figure \ref{fig:cno}. The [C/O] ratios in our yields show
  reasonable agreement with the CEMP observations at this
  metallicity. However the [C/N] ratios are too low compared to most of the
  CEMP values.

 We also present in Figure \ref{fig:cno} the computed yields under the effect of dilution in the
 secondary component's envelope.  We estimated
 the effects of dilution by calculating what would happen if 1 $\%$ of the
 material ejected from a primary star were accreted and homogeneously mixed
 in the 0.3 ${ M_{\odot}}$ convective envelope of a 0.8 ${
   M_{\odot}}$ secondary giant star.  This percentage of accretion is
 consistent with the value obtained using the Bondi-Hoyle approximation
 \citep{bon44}, as in \citet{hur02}, with a wind velocity of 10 ${
   km/s}$ characteristic of the stellar radii of our TP-(S)AGB stars, and
 the parameter $\alpha_w$ = 3/2. 

The approximation of mixing into a 0.3 ${ M_{\odot}}$ envelope is
reasonable in the case of giant secondary stars, whereas in the case of
dwarfs a more realistic approach to the problem should consider
the structure of the accreting star and an appropriate treatment of the
mixing mechanisms in their thin convective envelopes \citep{sta07}.
We find our
dilution estimates do not significantly alter the result of our comparisons
with the CEMP observations (Figure \ref{fig:cno}).

As mentioned above we have not yet computed nucleosynthesis with an
extended network and therefore we cannot compare s-process yields to
observations.  But given the high temperatures reached at the intershell
convective regions and the values of the dredge-up parameter for most of
our models (see Section 4), we can expect the presence of s-process
elements in the real counterparts of our model stars. This is also
consistent with the high fraction (about $\rm {80\%}$) of s-enriched CEMP
stars \citep{aok07}.

  \begin{figure}[t]
   \centering
   \includegraphics[scale=0.60,angle=0]{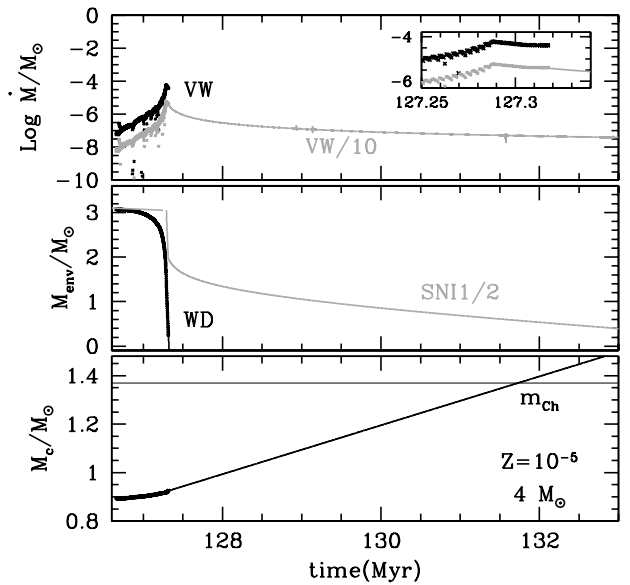}
   \caption{Variation of mass-loss rates, envelope mass and core mass versus time along the TP-AGB phase of the 4 ${ M_{\odot}}$ star.
The thick black lines correspond to the results of our full calculations and thereafter values
represented in thin lines correspond to extrapolations. Black lines correspond to VW mass-loss rates and grey lines correspond to the case VW/10 (see full text for details). The insert in the top panel highlights the evolution of the 
mass-loss rates at the end of our full computations.
The Chandrasekhar mass, ${ m_{Ch}}$, is shown for reference in the lower panel.}
              \label{fig:wind4}
    \end{figure}


The final fates of TP-(S)AGB stars is the result of a competition between
core growth and mass loss rates which, as we commented in Section 2,
are particularly uncertain at very low metallicities.  In 
Figure \ref{fig:mcores} we show the core masses after the SDU or DO
episodes and at the end of the computed TP-(S)AGB versus the initial masses
of our models.  It can be seen that the cores of our model stars do not
reach the Chandrasekhar mass and therefore these objects are able to avoid
becoming either SNe I$\rm{1/2}$ in the case of stars hosting CO cores
\citep{arn69}, or electron-capture supernovae \citep{nom84} in the case of
stars hosting ONe cores.  

   \begin{figure}[t] 
   \centering
   \includegraphics[scale=0.60,angle=0]{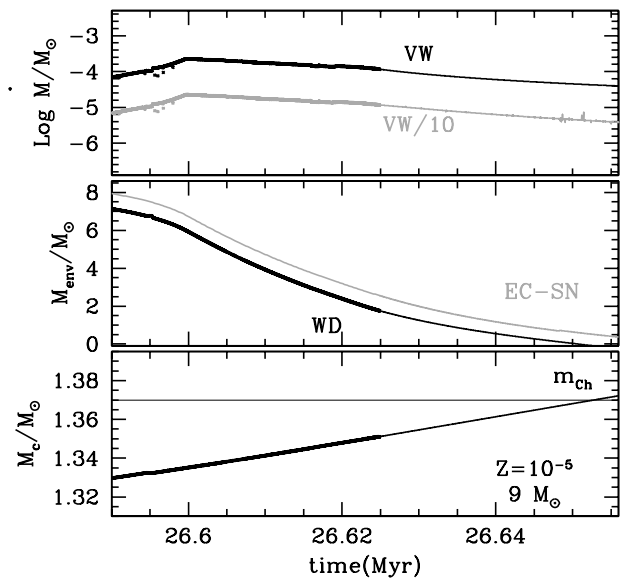} 
   \caption{Variation of mass-loss rates, envelope mass and core mass versus time along the TP-SAGB phase of the 9 ${ M_{\odot}}$ star.
The lines have the same meaning as in \ref{fig:wind4}.}
              \label{fig:wind9}
    \end{figure}

\begin{figure}[t]
   \centering
   \vspace{+0.0cm}
   \hspace{-0.3cm}
   \includegraphics[scale=0.55,angle=0]{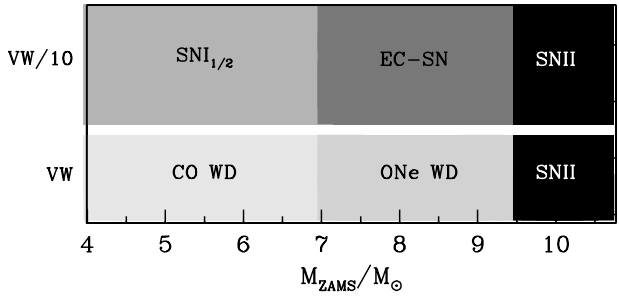}
   \caption{Mass limits at $Z=10^{-5}$ for the formation of carbon-oxygen and
            oxygen-neon white dwarfs,
            supernovae type ${\rm I1/2}$ (for stars whose CO core reach ${ m_{Ch}}$),
            electron-capture supernovae (for stars whose ONe cores reach ${ m_{Ch}}$) and supernovae type II.
            These mass limits are represented for the cases in which \citet{vwo93} are used (full calculations),
            and for the cases in which this rates
            are decreased one order of magnitude.
           }
              \label{fig:masslim}
    \end{figure}

It has recently been proposed by \citet{woo11} that stars of very low metallicities
may not experience strong enough pulsations that could lift envelope matter
to the photosphere and, by cooling, form grains. It is radiation pressure
onto these grains that drives stellar winds, so their absence might
switch-off or decrease substantially such winds. 
Mass loss rates are critical for the final fates of our model stars, so in
this section we briefly explore this uncertainty. As a test we consider how
the final fates of our models would change if the VW93 prescription is
overestimating the mass loss rates by one order of magnitude.  In this
case we did not follow the complete evolution but, instead, we 
divided the VW93 rates obtained along our full calculations by a factor 10,
starting at the beginning of the TP-(S)AGB phase. Then we integrated
these mass loss rates in time to obtain the corresponding envelope
masses. As one may expect the final envelope masses we get in this way are
still very high (more than 90\% of the original values). We use Bulirsch
and Stoer algorithm \citep{bur80} to extrapolate the mass loss rate values
beyond those that we can obtain directly. Once more these values can be
integrated to obtain the evolution of the envelope masses until the time
when the envelope has been completely ejected. We assumed that the
core growth rates are the same as the ones we find when using VW93.  Our
results for the 4 and the 9 ${ M_{\odot}}$ cases are show in in Figures
\ref{fig:wind4} and \ref{fig:wind9}, respectively. As we can see, for both
cases, the decrease of one order of magnitude in the standard rates is
enough to recover the possibility of the occurrence of
supernovae. If the initial mass is less than about 7 
${ M_{\odot}}$, our model stars would develop degenerate CO cores and
  explode as SNI${\rm 1/2}$, whereas if their initial mass is between 7
  ${ M_{\odot}}$ and 9 ${ M_{\odot}}$, they would develop ONe cores
  and undergo an electron-capture supernova.   Hence we can not
  confidently say that $Z = {\rm 10^{-5}}$ stars avoid the
  possibility of ending their lives as SNI${\rm 1/2}$. It is also
interesting to see that the extrapolations for the 9 ${ M_{\odot}}$
standard VW93 case show that the time for the core to reach ${ m_{Ch}}$
and the time necessary to complete the envelope ejection are very
similar. Therefore our 9 ${ M_{\odot}}$ case is very close to the limit
for the formation of EC-SNe at the end of its TP-SAGB phase. For a
comparison, the minimum initial mass for our models to reach ${ m_{Ch}}$
near the end of core carbon burning is $\sim$ 9.3 ${ M_{\odot}}$.  The
summary of our results for different initial masses and wind rate
prescriptions is shown in Figure \ref{fig:masslim}.

\section{Summary}
The evolution of model stars of metallicity $Z=10^{-5}$ and masses between
4 and 9 $M_{\odot}$ has been followed from the main sequence until the
Fe-opacity peak instability described in \citet{lau12} halts the
calculations. 
The very large increase in
metal abundances in the envelope caused by the dredge-out and
corrosive second dredge-up episodes gives rise to rather cool and extended
stellar envelopes during the early and TP-(S)AGB phases. This causes
mass loss rate values comparable to those of moderate-metallicity objects,
when keeping the prescription by \citet{vwo93} to characterise stellar
winds.

Our stars between 4 and 8 ${ M_{\odot}}$ experience the combined effects
of TDU and HBB along their TP-(S)AGB evolution.  The TDU parameter
$\lambda$ is near 0.9 for our lowest mass models and decreases to 0.5 and
0.05 (practically negligible) for our 8 and 9 ${ M_{\odot}}$ models,
respectively.  TDU is responsible for a steady increase in C surface
abundances in models between 4 and 8 ${ M_{\odot}}$ and, even though we
did not follow neutron-capture in this work, is also expected to enrich
in s-process elements the surfaces of these stars. All our models
experience a very strong HBB which results in significant surface enrichment in
N.

Given the uncertainty in the outcome of our model stars once the Fe-peak
instability is reached, we computed two sets of stellar yields based
on two different assumptions. In the first case we assumed that the
models can not recover after the instability and that the remnant envelopes
are lost on dynamical time scales. In the second case (synthetic treatment)
we assumed that the models recover after the instability and proceed
further along the TP-(S)AGB phase. We showed that the second treatment
has negligible effects on the lowest mass models, but it
gives C yields which are ${\rm 11\:\%}$, ${\rm 24\:\%}$ and ${\rm 22\:\%}$
higher for the 6, 7 and 8 ${ M_{\odot}}$ models respectively.  This is a
consequence of the TDU remaining operative during the last thermal
pulses. The 9 ${ M_{\odot}}$ model, in which TDU is negligible but HBB
is very important, yields ${\rm15\:\%}$ more N when the synthetic treatment
is used.

We also considered the uncertainties that surround mass loss at the
studied metallicity range. In particular we explored the possibility that
we might be overestimating mass-loss rates; see, for instance,
\citet{woo11}.  We artificially evolved our model stars along the
TP-(S)AGB considering mass-loss rates between the standard values and down
to one order of magnitude below. This decrease is enough to prevent the
fast ejection of the stellar envelope and, therefore, to allow the
formation of SNI$1/2$ in the case of stars of initial mass up to 7 ${
  M_{\odot}}$, or electron-capture SNe in the case of stars of initial
masses between 7 and 9 ${ M_{\odot}}$. Stars above this mass would end
up their lives as core-collapse supernovae.

It must be kept in mind that a reduction in mass loss rates due to stellar
winds is not the only reason why very low metallicity intermediate mass
stars might end their lives as SNeI$1/2$. As reported by \citet{lau08}, who
computed primordial intermediate-mass stars, these objects might have a
peculiar behaviour in the midst of their TP-(S)AGB. Their thermal pulses
gradually become weaker and eventually stop (while the model star keeps
numerically stable) when their envelopes still contain a few solar
masses. From this point on, the stellar cores keep growing up to the
Chandrasekhar mass and the model stars are expected to end their lives as
SNeI$1/2$.
In \citet{lau08} scenario thermal pulses cease because He-burning becomes stable
due to the high temperatures reached in the region where it is active. Such high temperatures
are favoured by the existence of a relatively massive degenerate core 
(1.1 $M_{\odot}$) and specially by the low envelope CNO mass fraction
(about $10^{-6}$, which allows very hot H-burning) 
and by the lack of occurrence of TDU (this process would cool 
the intershell region). In our $Z=10^{-5}$ models TDU is very efficient for
initial masses up to 7 $M_{\odot}$, so these stars seem likely to avoid the 
\citet{lau08} scenario, even if lower mass-loss rates allowed a longer 
TP-(S)AGB phase. On the other hand, our more massive models 
do not experience such efficient TDU episodes and might eventually develop 
stable He-shell burning in spite of their high envelope CNO mass fractions. 
Either way, this phenomenon has not been fully explored and, even though 
it does not seem likely to occur, we cannot confirm or discard its occurrence in our model stars. 

Most of the matter ejected by our stars is expected to enrich the ISM, but
a small fraction might be accreted by companion stars of wide binaries. We
 compared our results to observations of EMP and CEMP stars from the
SAGA database in order to see if our yields fit in the observed ranges. 
Our [C/O] yields are consistent with these observations,
  however the [C/N] yield values are too low compared to most of the observations of CEMP stars.

It would be interesting to obtain a wider set of yields with a
postprocessing nucleosynthesis code in order to perform further comparisons
with observational data in the near future.  Still, the huge number of
thermal pulses to follow and the strict spatial and temporal resolution
required makes this task enormously demanding.

\begin{acknowledgements}
      This work was supported by Jordi Ortiz Domenech and the Monash Centre for Astrophysics (MoCA).
      PGP would also like to thank the MoCA group for their kind
      hospitality and for the stimulating research atmosphere they provide.
      S.W.C. acknowledges support from the Australian Research Council's Discovery Projects funding scheme (project DP1095368) and a Monash Early Career Researcher grant (2012).
\end{acknowledgements}
\bibliographystyle{aa} 
\bibliography{aa2013_revised.bib}

\end{document}